\documentclass{aa} 
\usepackage{graphicx}
\usepackage{txfonts}
\usepackage{amsmath,systeme}
\usepackage{dcolumn}
\usepackage{bm}
\usepackage{csquotes}
\usepackage{graphicx}
\usepackage[english]{babel}
\usepackage[T1]{fontenc}
\usepackage{setspace}
\usepackage{geometry}
\usepackage{enumerate}
\usepackage{caption}
\usepackage{subcaption}
\usepackage{lineno}
\usepackage{algorithm}
\usepackage[]{algpseudocode}
\usepackage[dvipsnames]{xcolor}
\usepackage{hyperref}
\hypersetup{
    colorlinks,
    citecolor=blue,
    filecolor=blue,
    linkcolor=blue,
    urlcolor=blue
} 
\usepackage[normalem]{ulem}

\newcommand{\vmp}[1]{\textcolor{black}{#1}}

\def\be{\begin{equation}}
\def\ee{\end{equation}}
\def\bea{\begin{eqnarray}}
\def\eea{\end{eqnarray}}
\begin{document}

   \title{{Bayesian \vmp{t}est of Brans-Dicke \vmp{t}heories with \vmp{p}lanetary \vmp{e}phemerides: Investigating the \vmp{s}trong \vmp{e}quivalence \vmp{p}rinciple}}

\author{V. Mariani
        \inst{1}
        \and
        O. Minazzoli
        \inst{2,3}
        \and
        A. Fienga
        \inst{1,4}
        \and
        J. Laskar
        \inst{4}
        \and
        M. Gastineau
        \inst{4}
        }

   \institute{G\'eoazur, Universit\'e C\^ote d'Azur, Observatoire de la C\^ote d'Azur, CNRS, 250 av. A. Einstein, 06250 Valbonne, France\\
              \email{vincenzo.mariani@geoazur.unice.fr}
         \and
             {Universit\'e C\^ote d'Azur, Observatoire de la C\^ote d'Azur, CNRS, Artemis, bd. de l'Observatoire, 06304, Nice, France} \\
        \and 
             Bureau des Affaires Spatiales, 2 rue du Gabian, 98000  Monaco\\
        \and 
            IMCCE, Observatoire de Paris, PSL University, CNRS, 77 av. Denfert-Rochereau, 75014 Paris, France \\
             }

   \date{Received ...; accepted ...}

 
  \abstract

   {{We are testing the Brans-Dicke class of scalar tensor theories with planetary ephemerides}.}
   {In this work, we apply our recently proposed Bayesian methodology \vmp{to} the Brans-Dicke case, with an emphasis on the issue of the \vmp{strong equivalence principle (SEP)}.}
   {{We use \vmp{an} MCMC approach coupled to full consistent\vmp{,} planetary ephemeris construction (from point-mass body integration to observational fit) and} compare the posterior distributions obtained {with and without the introduction} of  potential violations of the \vmp{SEP}.}
   {We observe a shift in the confidence levels of the posteriors obtained.
   We interpret this shift as marginal evidence suggesting that the effect of \vmp{violation of the SEP} can no longer be assumed to be negligible in planetary ephemerides with the current data.   
   We also notably report that the constraint on the Brans-Dicke parameter with {planetary ephemerides} is getting closer to the figure reported from the Cassini spacecraft alone, but also to the constraints from {pulsars}. We anticipate that data from future spacecraft missions, such as BepiColombo, will significantly enhance the \vmp{constraints based on} planetary ephemerides.
   }
   {}


   \maketitle
%


\section{Introduction}

{Planetary ephemerides have evolved \vmp{over the years in line with improvements in} the observational accuracy \vmp{of} the astrometry of planets and natural satellites using the navigation tracking of these systems. The motion of the celestial bodies in our Solar System can be solved directly by integrating  their equations of motion \vmp{numerically}. The improved accuracy in measurements of observables from space missions (e.g. Cassini-Huygens, Mars Express, BepiColombo, Juice, etc.) makes the Solar System a suitable environment \vmp{for testing} general relativity theory (GRT) as well as alternative theories of gravity by means of Solar System ephemerides. {For a detailed review regarding the use of planetary ephemerides for testing gravity, we refer to \cite{fienga:2023lr}.}
Since 2003\vmp{,} the INPOP (Intégrateur Numérique Planeétaire de l'Observatoire de Paris)  planetary ephemerides \vmp{have been} developed \citep{Fienga2008,Fienga2009} with numerical integration of the Einstein-Infeld-Hoffmann equations of motion proposed by \cite{moyer2003monography}, and  \vmp{by} fitting the parameters of the dynamical model to the most accurate planetary observations following \cite{Soffel_2003}.}
In \vmp{the present} work, we use the INPOP21a planetary ephemerides \citep{INPOP21a}, \vmp{which} benefit from the latest Juno and Mars orbiter tracking data up to 2020, \vmp{and from} radio-science observations from space missions \vmp{such as} Cassini-Huygens, Mars Express, and Venus Express. Moreover, a fit of the Moon--Earth system to \vmp{Lunar Laser Ranging (LLR)} observations up to 2020 is included. The dynamical \vmp{modelling} used in INPOP21a takes into account the eight planets \vmp{of the Solar System, as well as} Pluto, the Moon, the asteroids, the Kuipier belt, the Sun oblateness, \vmp{and the Lense--Thirring effect.
We refer the reader} to \cite{INPOP21a} for a detailed review of this \vmp{version of INPOP}.
In \vmp{recent years}, the post-Newtonian (PPN) parameters have already been tested with planetary ephemerides \citep{fienga:2023lr}. However\vmp{,} previous derivations did not take into account the potential violation of the \vmp{strong equivalence principle} (SEP) owing to its assumed small impact on the ephemerides.
In other words, the effect of the gravitational self-energy of the planets violating the SEP was not implemented in the equations of motion. This decision was based on order-of-magnitude derivations \vmp{that indicated that} the effect was not deemed significant given the accuracy of the previous ephemerides.

Nevertheless, \vmp{almost} all alternative theories predict a violation of the SEP---that is, they predict that bodies with different self-gravitating energies fall differently in a given gravitational field \citep{nordtvedt1968pr1,nordtvedt1968pr2,will2018book}. In the PPN framework, this effect is parameterized by the Nordtvedt parameter $\eta$---see Eq. (\ref{eq:eta_alpha0})---in reference to \cite{nordtvedt1968pr1,nordtvedt1968pr2}, and $\eta=0$ in general relativity. Therefore, it was only a matter of time before planetary ephemerides reached a level of accuracy that necessitated \vmp{that this effect be accounted for} in the integration of the equations of motion.

In the present work, we implement the whole set of relevant equations in the Brans-Dicke case---including the effects of the predicted violation of the SEP. We use the recent work of \cite{bernus:2022pr} \vmp{, who derived} the whole phenomenology of Einstein-dilaton theories, and restrict our analysis to the one-parameter family that corresponds to Brans-Dicke theories. As we show in Sect. \ref{sec:graveqs}, one of the reasons for caution is that there is a potential ambiguity regarding the types of mass ({{{gravitational}}} or {{{inertial}}}) that appear in various equations, namely the equation of motion, the Shapiro delay, and the definition of the barycenter. We therefore start from a consistently derived phenomenology.

The motivation for investigating violations of the SEP in planetary ephemerides within the relatively limited framework of Brans-Dicke theories is two-fold. Firstly, as explained in a recent review \citep{fienga:2023lr}, \vmp{because of} the correlations between the parameters of the Solar System (masses, initial conditions, shape, etc.) and the theoretical parameters that parameterize deviations from the laws of general relativity (e.g. PPN parameters), it is somewhat difficult to come up with statistically robust tests of alternative theories to general relativity. \cite{fienga:2023lr} indeed notably argue that one should focus on each set of theories in a consistent manner, and this is our aim for the present study. Secondly, a robust statistical Bayesian methodology \vmp{was recently} developed and used \vmp{to test} a potential Yukawa suppression of the Newtonian potential in the Solar System \citep{marianietal2023}. It is interesting to apply this new methodology to other alternative phenomenologies. However, it is  feasible to apply such an approach \vmp{to phenomenologies that have just}  one theoretical parameter to be tested. \vmp{Therefore}, if the methodology is applied \vmp{to} scalar-tensor theories, one must restrict the study to the Brans-Dicke case.

\section{Dynamical and statistical formalism}

\subsection{Gravitational framework}
\label{sec:graveqs}

A discrepancy in the free fall of particles with differing compositions in a particular gravitational field would signify a violation of the weak equivalence principle (WEP). Meanwhile, a deviation for extended bodies with distinct gravitational self-energies would indicate a breach of the gravitational weak equivalence principle (GWEP). The latter forms a component of the SEP\vmp{, as delineated by} \cite{will2018book}. In all instances, through planetary ephemerides, our objective is to verify the universality of free fall (UFF), \vmp{or, in other words,} to confirm whether all bodies, regardless of their characteristics, experience identical rates of fall in a given gravitational field.

The violations of both the WEP and GWEP share a characteristic feature: at the Newtonian level, their equations of motion are expressed as follows \citep{bernus:2022pr,minazzoli:2016pr}:

\be
\boldsymbol{a}_{\mathrm{T}}=-\sum_{A \neq T} \mu_A \frac{\boldsymbol{r}_{A T}}{r_{A T}^{3}} \left(1+\delta_{\mathrm{T}}+\delta_{A T}\right), \label{eq:uffvgen}
\ee
where {$\boldsymbol{a}_{\mathrm{T}}$ is the {coordinate} acceleration of body $T$ (whereas the letter $A$ is used as index for the other bodies); $\boldsymbol{r}_{A T}$ is the relative position of body $T$ with respect to $A$; ${r}_{A T}=|\boldsymbol{r}_{A T}|$, with} $\mu_A= G \times m^{G}_A$ being the body $A$ gravitational parameter; and $\delta_T$ and $\delta_{AT}$ are coefficients that depend on the composition of the bodies $T$ and $A$ for the case of the violation of the WEP only. The most general post-Newtonian phenomenology for Einstein-dilaton theories \vmp{was derived by} \cite{bernus:2022pr}, and depends on several parameters that are related to the coupling between matter and gravity. Because the exploration of a high-dimensional theoretical space is very demanding in terms of computation time, a reduction of the number of parameters was made owing to the roughly similar compositions of the celestial bodies, and a simple rejection sampling \vmp{was applied to} the solutions of the ephemerides in order to estimate the constraints of the theoretical parameters.

In what follows, we use \vmp{an opposing strategy}: given the recent very effective Bayesian methodology that is \vmp{well suited to} the one-parameter family of theories, we reduce the general Einstein-dilaton action to the case of only one parameter, effectively recovering the Brans-Dicke theory, and we apply the Bayesian methodology to this case.

Following \cite{bernus:2022pr}, the equation of motion of the Brans-Dicke special case corresponds to
   \begin{align}
                        \bm {a}_T=&-\sum_{A\neq T} \frac{\mu_A}{r_{AT}^3}\bm r_{AT}\left(1+\delta_T\right)  -\sum_{A\neq T} \frac{\mu_A}{r_{AT}^3c^2}\bm r_{AT}\Bigg\{\gamma v_T^2 +(\gamma+1)v_A^2 \nonumber\\  
                         &-2(1+\gamma)\bm v_A.\bm v_T  -\frac{3}{2}\left(\frac{\bm r_{AT}.\bm v_A}{r_{AT}}\right)^2-\frac{1}{2}\bm r_{AT}.\bm a_A -2\gamma\sum_{B\neq T}\frac{\mu_B}{r_{TB}} \nonumber \\
    &+\sum_{B\neq A}\frac{\mu_B}{r_{AB}}\Bigg\} +\sum_{A\neq T}\frac{\mu_A}{c^2r_{AT}^3}\left[ 2(1+\gamma)\bm r_{AT}.\bm v_T -(1+2\gamma)\bm r_{AT}.\bm v_A \right] \nonumber \\
                &(\bm v_T-\bm v_A) + \frac{3+4 \gamma}{2}\sum_{A\neq T} \frac{\mu_A}{c^2r_{AT}}\bm a_A 
\, ,\label{eq:eq_eihmod}
                \end{align}

where {$\delta_T$ and the parameter $\gamma$ depend} only on a universal coupling constant $\alpha_0$, such that
\begin{equation}
  \gamma = \frac{(1-\alpha_0^2)}{(1+\alpha_0^2)} 
  \label{eq:defgamma}
\end{equation}
and 
\be
  \delta_T = - (1-\gamma) \Omega_T := \eta \Omega_T , 
  \label{eq:delta}
\ee
with
\be
 \Omega_T = \frac{3}{5} \frac{\mu_T}{c^{2} R_T}
,\ee
where $R_T$ \vmp{is} the radius of the planet T. \vmp{This} implies that the Nordtvedt parameter in this case is
\be
\eta = - (1-\gamma) = - 2 \frac{\alpha_0^2}{1+\alpha_0^2}.
\label{eq:eta_alpha0}
\ee

The Shapiro delay is also modified and is now such that the whole coordinate propagation time between an emitter and a receiver ($t_r-t_e$) reads as follows:
\begin{equation}
c(t_r-t_e)=R+\sum_T (1+\gamma-\delta_T)\frac{\mu_T}{c^2}\ln\frac{\bm{n}\cdot\bm{r}_{rT}+r_{rT}+\frac{4\mu_T}{c^2}}{\bm{n}\cdot\bm{r}_{eT}+r_{eT}+\frac{4\mu_T}{c^2}} ,
\label{eq:shapiro_dilaton}
                \end{equation}
  {where $R$ is the coordinate Euclidean distance between the emitter and the receiver}.
Another way to formulate the Shapiro delay is in terms of the inertial gravitational parameter,
\be
\mu_T^i = G \times m_T^i = (1-\delta_T) \mu_T,
\ee
where $m_T^i$ is the inertial mass of the body $T$, such that
\begin{equation}
c(t_r-t_e)=R+\sum_T (1+\gamma)\frac{\mu^i_T}{c^2}\ln\frac{\bm{n}\cdot\bm{r}_{rT}+r_{rT}+\frac{4\mu_T}{c^2}}{\bm{n}\cdot\bm{r}_{eT}+r_{eT}+\frac{4\mu_T}{c^2}} .
\label{eq:shapiro_dilaton2}
\end{equation}
In other words, the Shapiro delay is sensitive to the inertial masses, whereas the equation of motion depends on the gravitational masses.

Equally importantly, the definition of the Solar System barycenter (SSB) also involves the inertial gravitational parameter $\mu_A^i$. Indeed, from the Lagrangian formulation of the equations of motion, \cite{bernus:2022pr} show that the following barycenter constant vector is a first integral of the equations of motion, 
\begin{equation}
\bm{q}=\bm{G}-\bm{V}t ,\label{const_lag_dil}
\end{equation}
where
\begin{equation}
\bm{G}=\frac{c^2}{h}\sum_A\mu^i_A\bm{z}_A\left(1+\frac{v_A^2}{2c^2}-\frac{1}{2c^2}\sum_{B\ne A}\frac{\mu_B}{r_{AB}}\right) \label{bary_lag_dil}
\end{equation}
are the coordinates of the relativistic barycenter of the system and
\begin{equation}
\bm{V}=\frac{c^2\bm{P}}{h}
\end{equation}
is the velocity of the barycenter motion. $h$ is the conserved energy---whose value does not affect what follows but can be found in \cite{bernus:2022pr}---and $\bm{P}$ is the conserved linear momentum\vmp{, which reads}
\begin{align}
\bm{P}&=\sum_A\bm{p}_A \nonumber\\
&=\sum_A\mu^i_A\bm{v}_A\left[ 1 + \frac{1}{2c^2}\left( v_A^2 - \sum_{B\ne A}\frac{\mu_B}{r_{AB}} \right) \right] \nonumber\\
&\quad- \frac{1}{2c^2}\sum_A\sum_{B\ne A}\frac{\mu_A\mu_B}{r_{AB}}(\bm{n}_{AB}\cdot\bm{v}_A)\bm{n}_{AB}. \label{eq_momlin}
\end{align}

In this work, we \vmp{implemented} Eqs. (\ref{eq:eq_eihmod}) and (\ref{eq:shapiro_dilaton}) in the INPOP planetary ephemerides numerical integration of the planet equations of motion and its adjustment to observations, \vmp{whereas---because} the system has a conserved linear momentum---the equation for the SSB (defined with the inertial gravitational parameter) is integrated once at J2000, which is the {initial} date of integration for INPOP ephemerides. 
We  used INPOP21a  for this test, as in \cite{marianietal2023, mariani2023full}, by considering different values of $\alpha_0$ \vmp{and the changes this induces} in $\gamma$ and $\eta$. The obtained results are presented in Sect. \ref{sec:alpha0} in terms of $\gamma$, while Sect. \ref{sec:stat} explains the statistical method. 

To assess the effect of considering \vmp{the} potential violation of the GWEP (and \vmp{therefore} of the SEP) in planetary ephemerides, we constrained the value of $\gamma$ either assuming that $\eta=0$ (in Sect. \ref{sec:PPN})---as \vmp{was the case} in previous planetary ephemerides---or assuming that $\eta=-(1-\gamma)$ (in Sect. \ref{sec:alpha0}).

\subsection{{Statistical aspects}}
\label{sec:stat}
Starting from \cite{Fienga2015} ---see also in \cite{DiRuscioFienga2020} and \cite{Fienga2020P9}---, several tools \vmp{have been used} to determine the goodness of the INPOP fit with respect to modifications in the equations of motion or in the framework of the ephemeris, as well as the observations. Within this context, the computation of the INPOP $\chi^2$ plays a role \vmp{in determining} which model (or which data) \vmp{leads to a significant improvement of} the ephemeris computation. In \vmp{the present} work, the computation of $\chi^2$ is the output ---for a given value of $\gamma$--- of the INPOP iterative fit, \vmp{that is,} after the adjustment of all its astronomical parameters (see \cite{marianietal2023, mariani2023full, Bernus2019prl}). The $\chi^2(\gamma)$ is computed following \eqref{eq:chi2} 
\begin{equation}\label{eq:chi2}
\chi^2 \left( \gamma, \mathbf{k} \right) \equiv \frac{1}{N_{\text{obs}}} \sum_{i=1}^{N_{\text{obs}}} \left( \frac{g^i( \gamma, \mathbf{k}) - d^i_{\text{obs}}}{\sigma_i} \right)^2
,\end{equation}    
where $\gamma$ is a fixed value, $\mathbf{k}$ are the astronomical parameters fitted with INPOP, $N_{\text{obs}}$ is the number of observations, the function $g^i$ represents the computation of observables, the vector $\mathbf{d}_{\text{obs}}=(d^i_{\text{obs}})_i$ is the vector of observations and $\sigma_i$ are the observational uncertainties. 
The goal of the methodology applied is to obtain a posterior for {the parameter $\gamma$}. The general pipeline used to obtain such a posterior is similar to \vmp{that adopted by} \cite{marianietal2023} and \cite{mariani2023full}. We compute the value $\chi^2 (\gamma)$ (see Eq. \eqref{eq:chi2}) for several different values of $\gamma$, spreading over \vmp{our domain of} interest. 
For any given $\gamma$, the value $\chi^2(\gamma)$ is obtained as an outcome of the full INPOP iterative adjustment, \vmp{with $\gamma$  fixed}. In this way, the astronomical parameters are adjusted with the least-squares procedure, whereas $\gamma$ is a fixed value. 
Fixing the value of $\gamma$ is necessary \vmp{as} it ensures the $\gamma$ contribution to the dynamics, avoiding high correlations between $\gamma$ and the other astronomical parameters $\mathbf{k}$ (see \cite{Anderson197843, Bernus2019prl} for further details about correlation problem).
In order to produce the posterior, we use a Markov chain Monte Carlo (MCMC) method, the Metropolis--Hastings (MH) algorithm. In order to produce the MCMC, it is necessary to evaluate the likelihood  \vmp{sequentially, and therefore} the $\chi^2$, for thousands of different values of $\gamma$ (see \cite{mariani2023full} for further details).
\vmp{As} the computation of $\chi^2(\gamma)$ \vmp{is} expensive in terms of time, it becomes difficult to run the MCMC with the $\chi^2$ direct calculation. In order to overcome the problem of computation time, we apply a Gaussian process regression (GPR) to interpolate among the values $(\gamma, \chi^2(\gamma))$ already computed beforehand. 
Starting from this set of points, we obtain, thanks to the GPR, a function $\gamma \longmapsto \tilde{\chi}^2(\gamma)$, together with an uncertainty relative to the possible error of interpolation. The regression $\gamma \longmapsto \tilde{\chi}^2(\gamma)$ is necessary to run the MH algorithm, which has, as an outcome, the posterior density.
A prior density distribution \vmp{has to provided for the parameter} to sample with the MCMC. We used a uniform prior distribution for $\gamma$. 
The MH algorithm is based on the idea that the drawings from the posterior distribution are done sequentially and according to a random acceptance process. The outcome of the algorithm is a sequence of random samples from the posterior (this sequence is the Markov chain) and its equilibrium distribution is the posterior itself. 
At each step of the algorithm, one new element of the sequence is computed and proposed, \vmp{and is} accepted or rejected. The acceptance/rejection is random, according to a certain probability (\vmp{which changes} at each step) based on the last accepted element of the sequence as well as the likelihood of the last element and of the new candidate element.

\section{Results}

\subsection{Limit on $\gamma$ {without SEP} {violation}}
\label{sec:PPN}
\begin{figure}
        \includegraphics[width=\columnwidth]{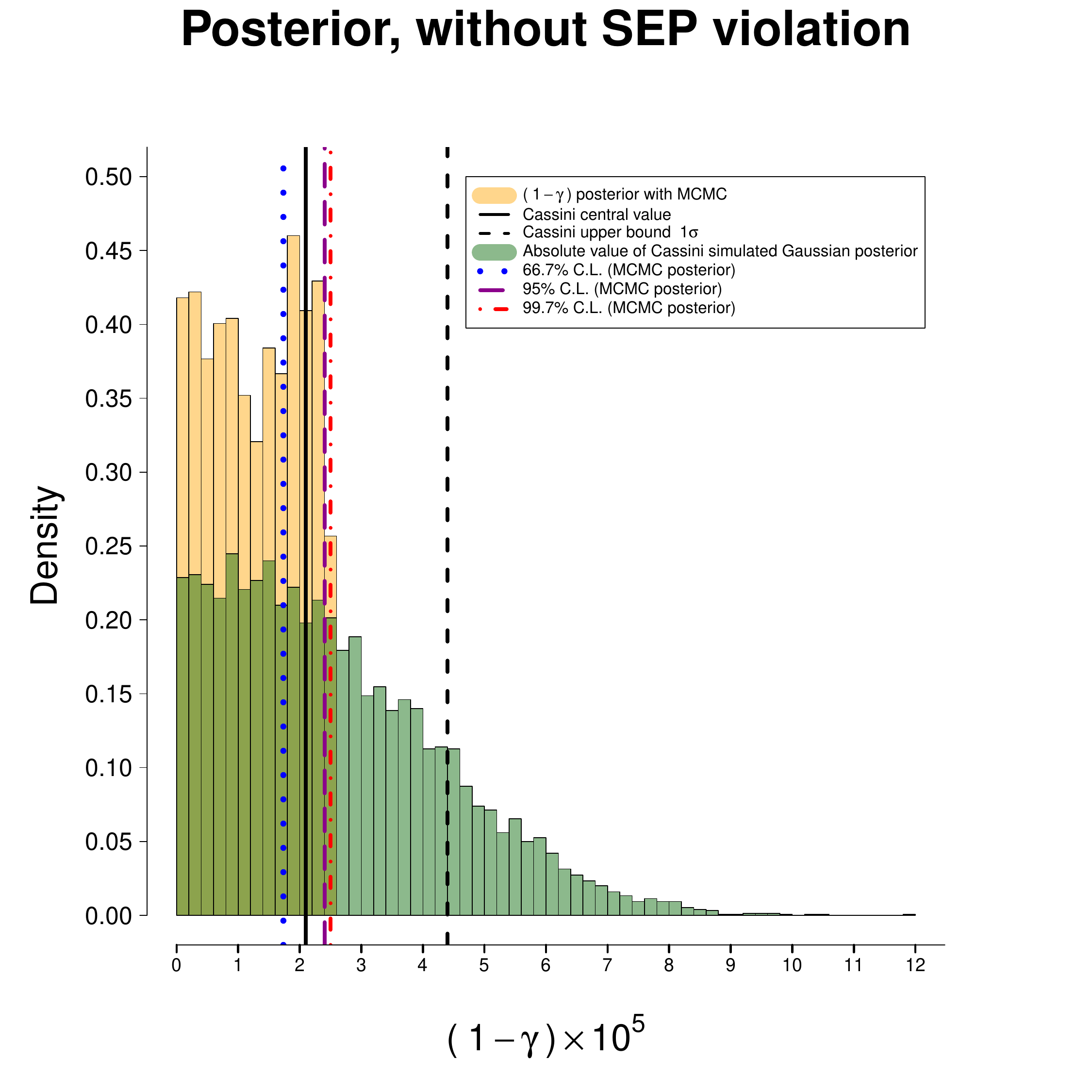}
    \caption{\vmp{Posterior distributions without SEP violation.} Yellow: {Posterior distribution for $1-\gamma$ without the implementation of the SEP violation.} Green: Absolute value of the simulated posterior for the determination of $1-\gamma$ with Cassini. Red line (dot-dashed): $99.7$ \% C.L. ($(1-\gamma)=2.50 \times 10^{-5}$). Blue line (dotted): $66.7$ \% C.L. ($(1-\gamma)=1.73 \times 10^{-5}$). Purple line (dashed): $95$ \% C.L. ($(1-\gamma)=2.40 \times 10^{-5}$). The prior for $(1-\gamma)$ is a uniform prior between $0$ and $15 \times 10^{-5}$. Solid black line: Absolute value of Cassini central value estimation ($(1-\gamma)=-2.1 \times 10^{-5}$). Dashed black line: Upper bound estimation from Cassini at $1\sigma$ C.L. }
    \label{fig:gamma_posterior_PPN}
\end{figure}

In order to evaluate the impact of the \cite{mariani2023full} method \vmp{within a well-known context}, we first \vmp{estimated the possible violation of general relativity} by considering $1-\gamma \ne 1$ without considering possible SEP violation (i.e. $\eta = 0$). In this case, the terms $\delta_T$ in Eq. (\ref{eq:eq_eihmod}) and $\delta_A$ in Eq. (\ref{eq:shapiro_dilaton}) vanish and only the terms with $\gamma$ remain.
With this configuration, our results are comparable \vmp{to} classical conjunction tests like the one obtained during the Cassini interplanetary phase when the radio-science signal of the Cassini s/c grazed the Sun before reception on earth. During this conjunction, the Shapiro delay is at its maximum and \cite{2003ASSL..293.....B} deduced the best constraint on $\gamma$ so far with a global fit of s/c orbit parameters and $\gamma$. As this part of the mission (interplanetary phase) is supposed to be the \vmp{least} affected by complex gravitational interactions or maneuvers, the determination is in principle the most accurate and the \vmp{least} affected by biases or noise. Because of the estimation method (least square), the \cite{2003ASSL..293.....B} constraint, $1-\gamma= (-2.1 \pm 2.3) \times 10^{-5}$, is given at 1$\sigma$, \vmp{assuming} a Gaussian distribution of the noise. 
Two main differences between our approach and the results obtained by \cite{2003ASSL..293.....B} have to be stressed. 
First, by construction of our model, only values of $\gamma$ \vmp{where} $1-\gamma > 0$ are tested. Once we compare our results with \cite{2003ASSL..293.....B}, we have to compare with the absolute value of their interval as it is represented in Fig. \ref{fig:gamma_posterior_PPN}. 
Second, we use a Bayesian approach using a uniform prior with a maximum value for $1-\gamma$ of $15 \times 10^{-5}$ and providing a posterior distribution of acceptable $1-\gamma$ values, as discussed in Sect. \ref{sec:stat}. In this context, for comparisons of results, \vmp{it is important to consider}  the confidence level (C.L.) deduced from the posteriors.
In Figure \ref{fig:gamma_posterior_PPN} and Table \ref{tab:res_eta}, we see the limit of about $4.4 \times 10^{-5}$ at $66.7 \%$ C.L. (1$\sigma$) of the $\|1-\gamma\|$ interval deduced from \cite{2003ASSL..293.....B} (black dashed line on Fig. \ref{fig:gamma_posterior_PPN}) and in yellow, the posterior distribution of $\|1-\gamma\|$ obtained with this work. At  $99.7 \%$ C.L., we obtain a limit of $2.5 \times 10^{-5}$ (red dot-dashed line on Fig. \ref{fig:gamma_posterior_PPN}) and of about $1.73 \times 10^{-5}$ at 66.7$\%$ C.L. (blue dotted line on Fig. \ref{fig:gamma_posterior_PPN}).
It is interesting to note, at this stage, that the limits we obtain are of the same order of magnitude \vmp{as those} obtained by  \cite{2003ASSL..293.....B}. \vmp{However, we stress} that these results were obtained \vmp{while} exploring only $\gamma$ values smaller than 1.

\subsection{{Limit on $\gamma$ and $\alpha_0$ with SEP violation}}
\label{sec:alpha0}

\begin{figure}
        \includegraphics[width=\columnwidth]{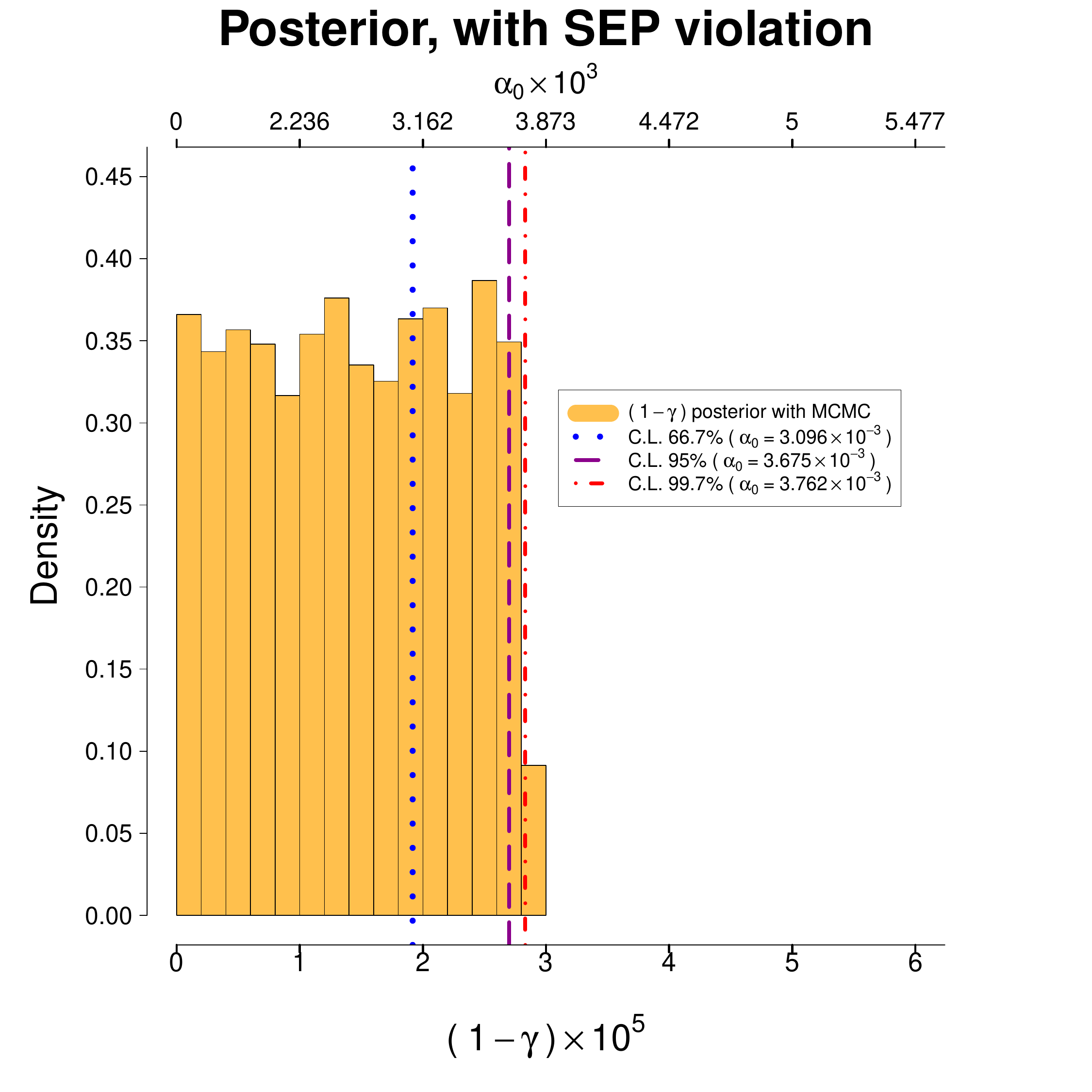}
    \caption{Posterior distribution for $1-\gamma$ and the corresponding values of $\alpha_0$ in the Brans-Dicke framework with the implementation of the SEP violation. Red line (dot-dashed): $99.7$ \% C.L. ($(1-\gamma)=2.83 \times 10^{-5}$ and $\alpha_0=3.762\times 10^{-3}$) . Blue line (dotted): $66.7 \%$ C.L. ($(1-\gamma)=1.92 \times 10^{-5}$ and $\alpha_0=3.096 \times 10^{-3}$). Purple line (dashed): $95$ \% C.L. ($(1-\gamma)=2.70 \times 10^{-5}$ and $\alpha_0=3.675\times 10^{-3}$). The prior for $(1-\gamma)$ is a uniform prior between $0$ and $15 \times 10^{-5}$.}
    \label{fig:alpha0_posterior}
\end{figure}

After considering the results of an exploration of the $1-\gamma$ {without SEP violation}, we now switch to the {full} Brans-Dicke formalism {with SEP violation} as described in Sect. \ref{sec:graveqs}. We used the method described in Sect. \ref{sec:stat} for the determination of acceptable values for the parameter $\alpha_0$. Fig. \ref{fig:alpha0_posterior}  \vmp{shows} the posterior of $1-\gamma$ and the deduced $\alpha_0$ from Eq. \ref{eq:defgamma}. At 99.7 $\%$ C.L., we obtain a limit on $\|1-\gamma \|$ of about $2.83 \times 10^{-5}$, leading to a constraint on $\alpha_0$ of about $3.76 \times 10^{-3}$.
{From} the constraint on $\alpha_0$ and on $\gamma$, one can deduce limits for the  Nordtvedt parameter $\eta=\gamma-1$. 
As given on Table \ref{tab:res_eta}, \vmp{at 99.7 $\%$ C.L.,  $\eta > -2.83 \times 10^{-5}$, at 95 $\%$ C.L., $\eta > -2.70 \times 10^{-5}$, and at 66.7 $\%$ C.L., $\eta > -1.92 \times 10^{-5}$}.
We stress that these constraints are only for $\eta < 0$ (see Eq.  \ref{eq:eta_alpha0}).
Different values of $\eta$ can be compared with these results {(see Table \ref{tab:res_eta})} but all are deduced from a direct fit   either of the s/c orbit, such as with \cite{genova:2018nc}, or of the Earth--Moon system \citep{2018MNRAS.tmp...86V}.

In \cite{genova:2018nc}, $\eta$ is obtained during the orbit determination of the s/c Messenger orbiting Mercury. In the global fit, \vmp{in addition to} orbit determination parameters, the PPN parameter $\beta$, the Sun oblateness $J_2$, and $\eta$ \vmp{are also estimated}. PPN $\gamma$ is fixed to the \cite{2003ASSL..293.....B} value. 

As discussed in \cite{fienga:2023lr}, the four parameters are strongly correlated, with or without the use of the Nordtvedt relation, 
\begin{equation}
\eta = 4 (\beta -1) - (\gamma -1) .
\label{eq:bad}
\end{equation}

For comparison, in Table \ref{tab:res_eta}, we \vmp{propose} some values found in the literature obtained with very different methods. In particular, we {consider} the thresholds obtained with \cite{bernus:2022pr} in the context of the Einstein-dilaton theories, which can be seen as a generalisation of the Brans-Dicke scalar--tensor theories. Indeed, in this case, coupling between matter and the \vmp{scalar field is} introduced and can take either a linear or a non-linear form. In \cite{bernus:2022pr}, limits on the universal coupling $\alpha_0$ have been obtained using planetary ephemerides (INPOP19a) together with constraints for the gaseous and the telluric planets, $\alpha_G$ and $\alpha_T,$ respectively. The comparison between the limit obtained for $\alpha_0$ in this work and \vmp{that from} \cite{bernus:2022pr} is not straight-forward, as in our case, we have $\alpha_G=\alpha_T=0$.

\section{Discussion}

\subsection{About the equivalence principle}

It is interesting to  note that the limit obtained for $1-\gamma$  in the BD framework with the SEP violation is  larger than \vmp{that} obtained without the SEP violation (see Fig. \ref{fig:gamma_posterior_PPN}). We interpret this as marginal evidence that, at the level of accuracy of current planetary ephemerides, the effect of the SEP violation is \vmp{no longer negligible. Indeed,} adding the effect increases the correlations between $\gamma$ and planetary parameters, which, here, comes from the additional correlations that exist between $\delta_T$ and planetary parameters (masses, initial conditions, etc.). This leads to the usual reduction in the accuracy of the estimation of the constraint on the theoretical parameter being tested. Indeed, \vmp{with a greater number of} correlations, more freedom is provided with which  to better fit the data; this leads to weaker constraints.
For a comprehensive discussion about the general issue of correlations between theoretical and planetary parameters, we refer the reader to \cite{fienga:2023lr}. Also, because we observe that including the SEP violation leads to slightly more stable solutions for the planetary orbits, \vmp{we believe} that including the SEP violation in alternative theories---that is, $\eta \neq 0$---is more consistent at the dynamical level than not \vmp{including it}.

\begin{table*}
    \centering
        \caption{Results and comparisons. {The first column gives the references of the published values where columns 2, 3, and 4 give the theoretical background, the method,  and the confidence level used for the publication, respectively. Finally, the last four columns give the obtained values corresponding to the following parameters: $\alpha_0$, $(1-\gamma)$, $\eta,$ and $\omega_{BD}$.}  }
\scalebox{0.9}{ 
    \begin{tabular}{l c c c c c c c}
    \hline
       Ref.  & Theory & Method & C.L. & $\alpha_0 \times 10^{3}$ &  $(1-\gamma) \times 10^{5}$ & $\eta \times 10^{5}$ &  $\omega_{BD}$\\
       \hline
       \\
      This work  & BD ($\eta = 0$) &  INPOP & 99.7 $\%$  &  & {$ \|1-\gamma \| < 2.50 $} &$  $ &  \\
      & & MCMC & 95 $\%$  & & {$ \|1-\gamma \| < 2.40 $} & & \\
      &&& {66.7} $\%$  & & {$ \|1-\gamma \| < 1.73 $} & & \\
      \\
      This work  & BD & {INPOP} & 99.7 $\%$  & {< 3.762} &  {$ \|1-\gamma \| < 2.83 $}  & $ \|\eta \| < 2.83 $ &  {$> 35,382$} \\
      & BD & MCMC & 95 $\%$  & {< 3.675} &  {$ \|1-\gamma \| < 2.70 $} & & \\
      &&& {66.7} $\%$  & {< 3.096} &  {$ \|1-\gamma \| < 1.92 $} & & \\
      \\
       \cite{bernus:2022pr} &  Dilaton & INPOP  & 99.7 $\%$ & $< 0.237$ & < 0.0112 & &  \\
      &&Cost function &&& & & \\ \\
\cite{2003ASSL..293.....B}& PPN & Cassini SC  & 1-$\sigma$& & $(-2.1 \pm 2.3)$ & & \\
& &&{66.7} $\%$  &&  $ \|1-\gamma \| < 4.4$ & & \\
       \cite{genova:2018nc} & PPN  wo Eq. \ref{eq:bad} & MSG  &1-$\sigma$ &&fixed  &$(-5.48 \pm 7.3)$ & \\
        & w Eq. \ref{eq:bad} &  &  1-$\sigma$&&fixed &$(-6.65 \pm 7.2) $ & \\
       \cite{viswanathan2018mn} & PPN & LLR & 3-$\sigma$ && 0 &$ \|\eta \| < 30 $ & \\
       \\
       \cite{2020voisin} & BD & {Pulsars} & 95$\%$  & <1.9 & < 0.76 & < 0.76 & > { $[1.3-1.4] \times 10^5$} \\
       \hline
    \end{tabular}
}
    \label{tab:res_eta}
\end{table*}

\subsection{{Conversion in terms of the usual Brans-Dicke parameter $\omega_{BD}$}}
\label{sec:BD}
{One can simply convert the constraint on $\alpha_0$ to a constraint on the usual Brans-Dicke parameter $\omega_{BD}$ through \citep{will:2014lr}
\begin{equation}
    \omega_{BD}= \frac{1}{2 \alpha_0^2}-\frac{3}{2}.
\end{equation}}
\vmp{Therefore}, based on the results given in Table \ref{tab:res_eta}, we can derive a 99.7$\%$ C.L. with $\omega_{BD} > 35,382$.
The limits obtained with the planetary ephemerides in this work are not as stringent as \vmp{those obtained by} \cite{2020voisin}. 
Indeed, using observations of a pulsar in a triple-star system, \cite{2020voisin} deduce that $\omega_{BD} > 130,000$---although this value depends on the equation of state being assumed for the neutron star, and \vmp{variation of a few tens of percent is generally found when using} different equations of state \citep{2020voisin}. Hence, constraints from planetary ephemerides are currently less powerful than the best constraints from pulsars. Nevertheless, we anticipate that future spacecraft missions, such as BepiColombo, will significantly improve the constraining power of planetary ephemerides in the near future.

\section{Conclusions}

In the present paper, we used a recent powerful Bayesian methodolgy developed for a one-parameter theoretical model in order to \vmp{obtain} the latest constraints from planetary ephemerides on the Brans-Dicke theory of gravity. In this context, we find that $\|1-\gamma \| < 1.92 \times {10^{-5}}$ at the $66.7 \%$ C.L., while the previous best Solar System constraints from ranging data of the Cassini spacecraft \cite{2003ASSL..293.....B} on $\gamma$ led to $\|1-\gamma \| < 4.4\times {10^{-5}}$ at $66.7 \%$ C.L. At the $99.7 \%$ C.L., the new planetary ephemerides constraint reads $\|1-\gamma \| < 2.83 \times {10^{-5}}$. While this is still inferior to the best constraints from pulsars \citep{2020voisin}, missions such as BepiColombo should improve this constraint in the near future.
We also \vmp{report} marginal evidence suggesting that the effect of the violation of the strong equivalence principle can no longer be assumed to be negligible with the current accuracy of planetary ephemerides.

\section*{Acknowledgements}

V.M. was funded by CNES (French Space Agency) and UCA EUR Spectrum doctoral fellowship. This work was supported by the French government, through the UCAJEDI Investments in the Future project managed by the National Research Agency (ANR) under reference number ANR-15-IDEX-01. The authors are grateful to the OPAL infrastructure and the Université Côte d’Azur Center for High-Performance Computing for providing resources and support. The authors thank G. Voisin for his useful inputs and discussions. This work was supported by the French group for gravitation and metrology PNGRAM (Programme National Gravitation, Références, Astronomie, Métrologie).



\bibliographystyle{aa}
\bibliography{Nordtvedt_in_ephemerides} 

\begin{thebibliography}{24}
\expandafter\ifx\csname natexlab\endcsname\relax\def\natexlab#1{#1}\fi

\bibitem[{Anderson {et~al.}(1978)Anderson, Keesey, Lau, Standish, \&
  Newhall}]{Anderson197843}
Anderson, J., Keesey, M., Lau, E., Standish, E., \& Newhall, X. 1978, Acta
  Astronautica, 5, 43

\bibitem[{Bernus {et~al.}(2019)Bernus, Minazzoli, Fienga, Gastineau, Laskar, \&
  Deram}]{Bernus2019prl}
Bernus, L., Minazzoli, O., Fienga, A., {et~al.} 2019, Phys. Rev. Lett., 123,
  161103

\bibitem[{{Bernus} {et~al.}(2022){Bernus}, {Minazzoli}, {Fienga}, {Hees},
  {Gastineau}, {Laskar}, {Deram}, \& {Di Ruscio}}]{bernus:2022pr}
{Bernus}, L., {Minazzoli}, O., {Fienga}, A., {et~al.} 2022, \prd, 105, 044057

\bibitem[{{Bertotti} {et~al.}(2003){Bertotti}, {Farinella}, \&
  {Vokrouhlick}}]{2003ASSL..293.....B}
{Bertotti}, B., {Farinella}, P., \& {Vokrouhlick}, D. 2003, {ASSL, volume 293},
  Vol. 293, {Physics of the Solar System - Dynamics and Evolution, Space
  Physics, and Spacetime Structure.}

\bibitem[{Di~Ruscio {et~al.}(2020)Di~Ruscio, Fienga, Durante, Iess, Laskar, \&
  Gastineau}]{DiRuscioFienga2020}
Di~Ruscio, A., Fienga, A., Durante, D., {et~al.} 2020, A\&A, 640, A7

\bibitem[{{Fienga} {et~al.}(2021){Fienga}, {Deram}, {Di Ruscio}, {Viswanathan},
  {Camargo}, {Bernus}, {Gastineau}, \& {Laskar}}]{INPOP21a}
{Fienga}, A., {Deram}, P., {Di Ruscio}, A., {et~al.} 2021, Notes Scientifiques
  et Techniques de l'Institut de Mecanique Celeste, 110

\bibitem[{Fienga {et~al.}(2020)Fienga, Di~Ruscio, Bernus, Deram, Durante,
  Laskar, \& Iess}]{Fienga2020P9}
Fienga, A., Di~Ruscio, A., Bernus, L., {et~al.} 2020, A\&A, 640, A6

\bibitem[{{Fienga} {et~al.}(2015){Fienga}, {Laskar}, {Exertier}, {Manche}, \&
  {Gastineau}}]{Fienga2015}
{Fienga}, A., {Laskar}, J., {Exertier}, P., {Manche}, H., \& {Gastineau}, M.
  2015, Celestial Mechanics and Dynamical Astronomy, 123, 325

\bibitem[{{Fienga} \& {Minazzoli}(2023)}]{fienga:2023lr}
{Fienga}, A. \& {Minazzoli}, O. 2023, arXiv e-prints (accepted in Living
  Reviews in Relativity), arXiv:2303.01821

\bibitem[{{Fienga, A.} {et~al.}(2009){Fienga, A.}, {Laskar, J.}, {Morley, T.},
  {Manche, H.}, {Kuchynka, P.}, {Le Poncin-Lafitte, C.}, {Budnik, F.},
  {Gastineau, M.}, \& {Somenzi, L.}}]{Fienga2009}
{Fienga, A.}, {Laskar, J.}, {Morley, T.}, {et~al.} 2009, A\&A, 507, 1675

\bibitem[{{Fienga, A.} {et~al.}(2008){Fienga, A.}, {Manche, H.}, {Laskar, J.},
  \& {Gastineau, M.}}]{Fienga2008}
{Fienga, A.}, {Manche, H.}, {Laskar, J.}, \& {Gastineau, M.} 2008, A\&A, 477,
  315

\bibitem[{{Genova} {et~al.}(2018){Genova}, {Mazarico}, {Goossens}, {Lemoine},
  {Neumann}, {Smith}, \& {Zuber}}]{genova:2018nc}
{Genova}, A., {Mazarico}, E., {Goossens}, S., {et~al.} 2018, Nature
  Communications, 9, 289

\bibitem[{Mariani {et~al.}(2023{\natexlab{a}})Mariani, Fienga, \&
  Minazzoli}]{mariani2023full}
Mariani, V., Fienga, A., \& Minazzoli, O. 2023{\natexlab{a}}, Testing the mass
  of the graviton with Bayesian planetary numerical ephemerides B-INPOP

\bibitem[{Mariani {et~al.}(2023{\natexlab{b}})Mariani, Fienga, Minazzoli,
  Gastineau, \& Laskar}]{marianietal2023}
Mariani, V., Fienga, A., Minazzoli, O., Gastineau, M., \& Laskar, J.
  2023{\natexlab{b}}, Phys. Rev. D, 108, 024047

\bibitem[{{Minazzoli} \& {Hees}(2016)}]{minazzoli:2016pr}
{Minazzoli}, O. \& {Hees}, A. 2016, \prd, 94, 064038

\bibitem[{Moyer(2003)}]{moyer2003monography}
Moyer, T. 2003, Formulation for Observed and Computed Values of Deep Space
  Network Data Types for Navigation (JPL Deep-Space Communications and
  Navigation Series), 1st edn.

\bibitem[{Nordtvedt(1968{\natexlab{a}})}]{nordtvedt1968pr1}
Nordtvedt, K. 1968{\natexlab{a}}, Phys. Rev., 169, 1014

\bibitem[{Nordtvedt(1968{\natexlab{b}})}]{nordtvedt1968pr2}
Nordtvedt, K. 1968{\natexlab{b}}, Phys. Rev., 169, 1017

\bibitem[{Soffel {et~al.}(2003)Soffel, Klioner, Petit, Wolf, Kopeikin,
  Bretagnon, Brumberg, Capitaine, Damour, Fukushima, Guinot, Huang, Lindegren,
  Ma, Nordtvedt, Ries, Seidelmann, Vokrouhlick, Will, \& Xu}]{Soffel_2003}
Soffel, M., Klioner, S.~A., Petit, G., {et~al.} 2003, The Astronomical Journal,
  126, 2687

\bibitem[{{Viswanathan} {et~al.}(2018{\natexlab{a}}){Viswanathan}, {Fienga},
  {Minazzoli}, {Bernus}, {Laskar}, \& {Gastineau}}]{2018MNRAS.tmp...86V}
{Viswanathan}, V., {Fienga}, A., {Minazzoli}, O., {et~al.} 2018{\natexlab{a}},
  \mnras [\eprint[arXiv]{1710.09167}]

\bibitem[{{Viswanathan} {et~al.}(2018{\natexlab{b}}){Viswanathan}, {Fienga},
  {Minazzoli}, {Bernus}, {Laskar}, \& {Gastineau}}]{viswanathan2018mn}
{Viswanathan}, V., {Fienga}, A., {Minazzoli}, O., {et~al.} 2018{\natexlab{b}},
  \mnras, 476, 1877

\bibitem[{{Voisin} {et~al.}(2020){Voisin}, {Cognard}, {Freire}, {Wex},
  {Guillemot}, {Desvignes}, {Kramer}, \& {Theureau}}]{2020voisin}
{Voisin}, G., {Cognard}, I., {Freire}, P. C.~C., {et~al.} 2020, A\&A, 638, A24

\bibitem[{{Will}(2014)}]{will:2014lr}
{Will}, C.~M. 2014, Living Reviews in Relativity, 17, 4

\bibitem[{Will(2018)}]{will2018book}
Will, C.~M. 2018, Theory and Experiment in Gravitational Physics, 2nd edn.
  (Cambridge University Press)

\end{thebibliography}






\label{lastpage}
\end{document}